\documentclass[aip,reprint,nofootinbib]{revtex4-1}

\usepackage{hyperref}
\usepackage{amsfonts}
\usepackage{amsmath}
\usepackage{graphicx}
\usepackage[utf8]{inputenc}
\usepackage{amssymb}
\usepackage{times}
\usepackage{color}

\def\>{\rangle}
\def\<{\langle}

\newcommand{\ket}[1]{|#1\rangle}

\newcommand{\fket}[1]{|#1\!\succ\!}

\newcommand{\tket}[1]{|\tilde{#1}\rangle}

\newcommand{\tfket}[1]{|\widetilde{#1}\!\succ\!}

\graphicspath{{originalfigs/}{spreadfigs/}} 

\begin{document}

\title{Photonic quantum computing with probabilistic single photon sources but without coherent switches} 

\author{Terry Rudolph}
\affiliation{Dept. of Physics, Imperial College London, London SW7 2AZ }
\email{t.rudolph@imperial.ac.uk}
\affiliation{PsiQuantum, Palo Alto.}

\date{\today}

\begin{abstract}
We present photonic quantum computing architectures that can deal with both probabilistic (heralded) generation of single photons and probabilistic gates without making use of coherent switching. The only required dynamical element is the controllable absorption of all photons in a given mode. While the architectures in theory scale polynomially in the resources required for universal quantum computation, as presented their overhead is large and they are illustrative extreme points in the configuration space of photonic approaches, rather than a recipe that anybody should seriously pursue. They do, however, prove that many things presumed necessary for photonic quantum computing, in fact are not. Of potentially independent interest may be that the architectures make use of qubits which have many possible microstates corresponding to a single effective qubit state, and the technique for dealing with probabilistic operations is to, when necessary, just enlarge the set of such microstates to incorporate all possibilities, while making heavy use of the subsequent ability to ``coherently erase'' which particular microstate a given qubit is in. 

\end{abstract}


\maketitle 

\emph{This paper is prepared for the Jonathan P. Dowling memorial issue of AVS Quantum Science. The last scientific discussion I had with Jon, circa mid-2017, he had read Ref.~\onlinecite{tezoptimistic} and was pressing me on aspects of linear optical quantum computing and silicon photonics. I had recently realized that architectures such as the ones presented herein were also possible, and I tried to tell him ``everything you think you know about what is strictly necessary for LOQC is wrong!'' Unfortunately at that time I couldn't tell him in detail about these results, and the chance never came.}

\section{Preamble}
 
This article contains the concatenation of two papers written between 2016-18 but never published. Other than minor editing and replacement of the introduction and conclusions by Sec.~\ref{sec:overview}, they are unchanged. This means the underlying assumptions about how to prove an architecture universal are based on slightly overcomplicated ideas regarding generation of cluster states and percolation theory and so on. Since that time we have realized that rather than overcoming the non-determinism of photonic gates by teleportation through implausibly large ancillae states\cite{Knill2001}, or via cluster state computation\cite{Raussendorf2001} using percolated graph states\cite{Browne2005,Kieling2007,Gimeno-Segovia2015,Zaidi2015} (which accumulate error as they are renormalized), it is possible to do \emph{fusion based quantum computing}\cite{bartolucci2021fusionbased} (FBQC) where we use a high erasure-threshold code to directly deal with \emph{both} the non-determinism of the gates and all error correction simultaneously. The exposition of the results in this article could have been a little clearer if FBQC had been the stated target from the start. 

\begin{figure}
\centering
\includegraphics[scale=0.35, angle=-90, clip=true,bb=3cm 3.5cm 18cm 29cm]{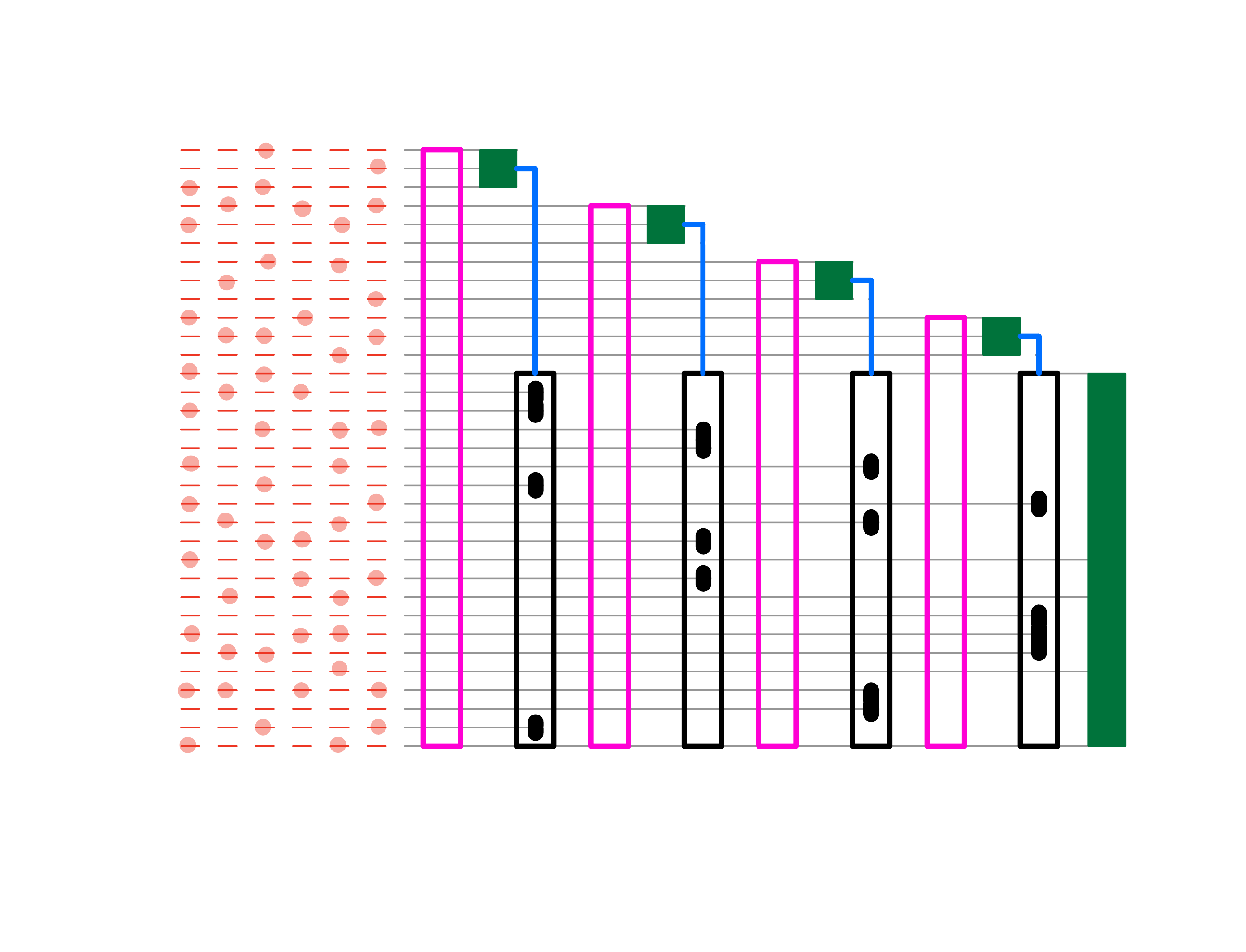}
\caption{An extremely high level overview of the first of the multi-rail architectures. Grey lines denote spatial modes. Spatial modes/time bins are occupied by (heralded) random single photons (red). These pass through alternating sequences of passive interferometers (pink), detectors (green) which via classical feedforward (blue) control subsets of the modes to be blocked (black). Only the configurations of modes being blocked depend on the initial random state of the photons and of the computation being performed, the passive circuits (which may contain fixed length delays) are predetermined and independent of both. Note also that larger/longer computations only (polynomially) increases the number of modes and time bins, the optical depth (number of  passive interferometers) experienced by any photon is independent of such.}
\label{fig:mroverview}
\end{figure}

\section{Overview of the architectures}\label{sec:overview}

It is typically presumed that high-efficiency sources of single photons will be necessary to do photonic quantum computing\footnote{Here we consider only finite-dimensional photonic quantum information.}. Roughly speaking there are two approaches to the construction of such: generate the photons from a deterministic emitter (ions, quantum dots etc), or generate pairs of photons randomly using a weak nonlinear-optical process, and herald the existence of the desired single photon by the detection of its partner. While the efficiency of the latter procedure may be low, considerations of achievable repetition rates, photon purity and compatibility with silicon photonics have led to proposals to ameliorate the low efficiency by \emph{multiplexing}\cite{Migdall2002}. That is, we build many copies of the heralded-but-inefficient-source, and then use a switch to select out a success. The same principle can then be applied at subsequent stages wherein one probabilistically creates larger and larger entangled states, until achieving a size suitable for use in a computation. 

Multiplexing in the fashion just described requires switches that maintain the quantum state of the light being switched. The switches used by the telecom industry to switch coherent (laser) light are also of this form, and from now on they will be referred to as \emph{coherent switches}. 

The architectures we present in this article investigate a different approach that lets us use randomly generated photons directly and yet requires no coherent switching! The approach still requires a dynamically controlled `switch', but it need only be of a ``blocking'' or ``incoherent/filtering/absorbing'' type\footnote{In certain technologies the controllable absorption-of/transparency-to light could be combined with the quantum Zeno effect to make an approximately coherent switch. That is not what we will be doing here.}. This is achieved at the expense of building much larger (but fixed, passive) interferometers. The methods presented should be viewed very much as ``proofs of principle'' - although in theory they scale polynomially with the size of the computations under consideration, the scaling of the versions presented here is not practical.  

A very high level summary of the first architecture is given in Fig.~\ref{fig:mroverview}.
 Note that all reasonable photonic architectures must have an optical depth which is constant with the size of the computation - that is any individual photon passes through only a constant (and hopefully $< 10$) number of components, because loss will accumulate exponentially if not. The architectures presented here satisfy this requirement.  

Even for the non-connoisseur of photonic quantum computing there are aspects of these results which may have more general interest. For example, they make use of qubits that have many possible (orthogonal) microstates corresponding to a single qubit state. While this sounds similar to an error correcting code, the mechanism here still only uses a single physical photon per qubit. The advantage of being able to associate many microstates with a single qubit state is that, when limited to probabilistic processes, we can lump together \emph{all} possible microstates output from repeated copies of the probabilistic process and redefine our qubit to encompass them all.       

Critical to getting away with this while still having our photons act as qubits is being able to perform \emph{coherent erasure} of microstate information, by which we mean a process wherein it is possible to destroy information regarding the specific microstate the system is in, while maintaining relevant phase information `between' the microstates encoding different possible qubit states\footnote{A simple example of `coherent erasure' from standard quantum information  would be destroying information about the computational basis value $\ket{0}/\ket{1}$ by performing a measurement in the $\ket{\pm}$ basis. This maintains phase relationships between systems that the erased qubit may be entangled with.}. 

%
%

\section{PAPER I: Architecture for photonic quantum computing based on multirail encodings and passive multiplexing}
\subsection{Stochastic Sources}

In what follows we will only consider photons which occupy discrete spatiotemporal modes. That is, a photon can only be found in one of a finite set of discrete waveguides, which define spatial modes $\vec{k_i}$, within a discrete time bin $t_j$. The degree of temporal discretization is typically provided by the mechanism responsible for generating the photons, the full temporal extent of the photon's wavepacket is presumed to fit within the time bin.      

One of our ultimate goals will be to show that even in the absence of coherent switches, rather than getting a photon into a single such spatiotemporal mode with high efficiency (``good single photon sources''), for universal photonic quantum computing it suffices to only get a single photon randomly into \emph{any one} of a set of spatiotemporal modes with high efficiency. The remaining modes within the set must be unoccupied (vacuum). Such a source will be termed a \emph{stochastic source}, although it should be emphasized that the stochasticity refers to the fact that each use of the source will produce a single photon in a random (though known) spatiotemporal mode, the \emph{efficiency}  - ie the probability that one (and only one) photon is produced - will need to still be close to 1.  

Consider then some variations on how such a source might be constructed. Imagine we have one (heralded) probabilistic single-photon source with efficiency $\eta$. By firing it $m$ times, we could get lucky and one and only one photon be produced. This would happen with probability $m\eta(1-\eta)^{m-1}$, which is an improvement in efficiency over the single source alone, although there is no way to make this probability arbitrarily close to 1.  We could, however, arrange to fire the source $m$ times, but turn off the pump as soon as a single photon has been produced. This `dump-the-pump' style switching has the advantage it does not involve directly switching the photons that will be our ultimate carriers of quantum information. The source efficiency would now be $1-(1-\eta)^m$, which goes to 1 for large $m$. A final alternative, achieving the same efficiency, is to fire the source $m$ times, and if more than one photon is produced to filter the extra photons out, using an incoherent `blocking' or `absorbing' style of switch. 

The $m$-mode source of the previous paragraph, however created, will be a ``purely temporal'' stochastic single-photon source, since the $m$ modes are distinguished only by time - the photons all lie in the same waveguide (spatial mode). We can also consider constructing a high efficiency ``purely spatial'' stochastic source, that is one where all $m$ modes are defined by multiple waveguides at a single time. For this case we could begin with $m$ heralded single photon sources firing into different waveguides.  By firing some or all of them simultaneously, potentially also with dump-the-pump and/or blocking style switching (and suitable fixed delays), a wide variety of possibilities exist to make purely spatial stochastic sources with high efficiency. 

The most general stochastic source would involve a mix of spatial and temporal modes. Whilst we might prefer to (say) use a purely spatial stochastic source for the single photons, at later stages in the architecture we may prefer to end up with our quantum information encoded in a mixed spatiotemporal form, and so will consider how to deal with this as well.       

To capture the various types of sources in a compact notation, we will denote by an $[a,b;p]$-source one that with probability $p$ has a single photon in any one of $a$ spatial modes and $b$ temporal modes (and vacuum in all other modes). For example, a regular heralded source at $20\%$ efficiency is a $[1,1;0.2]$-source. By pumping it 16 times (and turning off the pump or filtering appropriately) we create a purely temporal $[1,16;0.972]$-source, while by taking four such sources and pumping each up to 8 times we can create a mixed spatio-temporal $[4,8;0.999]$-source.

\subsection{Multi-Rail encoding of qubits}

A common way of encoding a photonic qubit is to pick any two modes of the electromagnetic field which differ only in a degree of freedom that can be manipulated passively with linear optics. For example, we might use polarization, or angular momentum or spatial mode, for definiteness we will stick to using the latter unless otherwise indicated\footnote{Encoding in frequency is fully compatible with the architectures - and in fact can make them considerable more spatially compact. Note that doing the coherent erasure (Hadmard) transformations requires a ``frequency beamsplitter'', a dynamical object\cite{frequencyreview}. Here the emphasis is on avoiding all dynamical objects other than blocking-type switches.}. The logical\footnote{``Logical'' in this paper never refers to a logical qubit from the theory of fault tolerance. It refers to the `bare' qubit for which a single qubit basis state might correspond to many physically-distinct states. Fault tolerant logical qubits need to be built on top of this architecture, e.g. via FBQC\cite{bartolucci2021fusionbased}} qubit basis $\ket{0},\ket{1}$ is then defined across the two chosen modes via
\begin{equation}
	\ket{0}\Leftrightarrow \fket{1}\fket{0}, \;\;\;\; \ket{1}\Leftrightarrow \fket{0}\fket{1},
\end{equation} 
where $\fket{n}$ denotes a single-mode Fock (number) state occupied by $n$ photons. This is often termed a `dual-rail' encoding.

By increasing the number of modes one can certainly encode higher-dimensional (i.e. qudit) quantum information, although there do not appear to be large advantages to doing so \cite{Joo2007,jakehigherd}. Here we instead introduce a somewhat odd encoding, where we take $2m$ discretized spatio-temporal modes, and the logical qubit state $\ket{0}$ is defined by allowing the photon to be in \emph{any one} of the first $m$ modes, while the logical state $\ket{1}$ is defined by allowing the photon to be in any one of the second $m$ modes. For example:  
\begin{eqnarray}\label{eq:multirailencodeI}
	\ket{0}&\Leftrightarrow  & \fket{0,0,1,0}\fket{0,0,0,0}\,\,=:\fket{1_3}\fket{0} \nonumber \\ 
	\ket{1}&\Leftrightarrow  & \fket{0,0,0,0}\fket{0,1,0,0}\,\,=:\fket{0}\fket{1_2}.
\end{eqnarray} 
The stochastic sources of the preceding section are then interpreted as a computational basis state preparation of a multi-rail qubit. 

For any given photon/qubit we will know which of the modes encode its computational basis, but, as the above example emphasizes, it is \emph{not} the case that the $\ket{0}$, $\ket{1}$ states for a single qubit need be defined using ``matching'' (i.e. complementary) pairs of modes $(i,m+i)$. This immediately raises the question, how can we (passively) do a single qubit gate? In the dual rail encoding, where there are only two (say) spatial modes, they are automatically ``matched appropriately'' (i.e. $i=m=1$ the only option) and so combining the corresponding modes on a 2-mode beamsplitter performs the desired transformation whilst automatically remaining in the subspace defining the qubit. In the example of Eq.~(\ref{eq:multirailencodeI}), a beamsplitter between modes 3 and 5 would be fine, but one between any other pairs of modes would take us out of the defined computational basis - for example giving terms that have both photons somewhere in modes $1,\ldots4$ or both in $5,\ldots8$, which are not multirail qubit states as defined. If we have to adapt the particular pair of modes we apply an operation to based upon which of the modes our (stochastic) single photon source happens to occupy, we would be back to requiring coherent switching. Fortunately, as we will see later, this will not be necessary. 

Moreover, if the modes we are using span several time bins then we may need a 'temporal beamsplitter' of some form, a device that cannot be constructed with purely passive linear optics. Again, fortunately this  will turn out to not be necessary.

\subsection{Production of maximally entangled states} 

In this section we restrict attention to production of multirail qubits from purely spatial stochastic sources.

\subsubsection{Review: Production of dual-rail encoded Bell pairs using deterministic sources} 

\begin{figure}
\centering
\includegraphics[scale=0.8, clip=true]{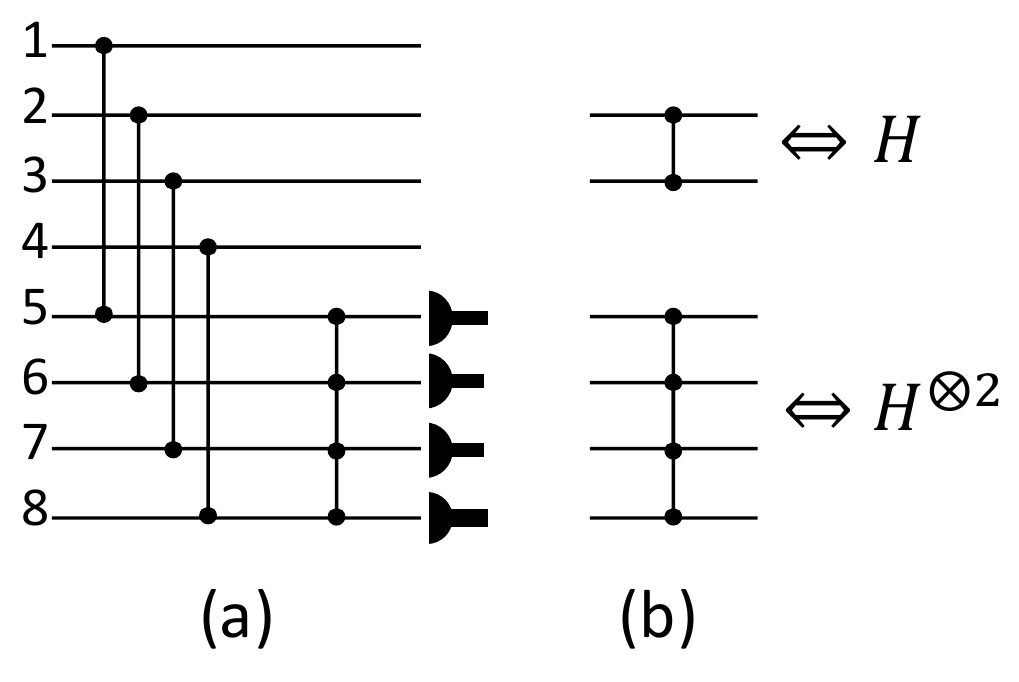}
\caption{(a) An 8-port interferometer that can generate maximally entangled dual-rail photonic qubits. Horizontal lines are modes (not qubits!), the vertical lines connecting modes indicate interference at a beamsplitter. For this interferometer, inputting a single photon to each of ports 1-4, combined with detection of single photons at any two distinct modes 5-8, produces a photonic Bell state on the output modes 1-4. (b) The notation used for generalized beamsplitters throughout this paper. A standard, 2-mode, 50:50 beamsplitter is described by the transformation (on mode operators) $H=\tfrac{1}{\sqrt{2}}\begin{bmatrix}1 & 1 \\ 1 & -1\end{bmatrix}$. On a $2^m$ port interferometer the notation indicates an interferometer described by the mode transformation $H^{\otimes m}$, which has a simple recursive decomposition (see Fig.~\ref{fig:bsrecursion})) in terms of $2m$ two-mode beamsplitters.}
\label{fig:bellcircuit}
\end{figure}

Recall a scheme for heralded generation of an entangled state \cite{Zhang2008} from four single photons. Fig.~\ref{fig:bellcircuit}(a) shows an interferometer with 8 spatial modes, that can take in four single photons (all in the same time bin) in spatial modes 1-4, and produce a maximally entangled pair of photons in the output spatial modes 1-4. Despite the apparent simplicity, the output state just prior to detection is a superposition of 144 terms - that is, a large amount of multiphoton interference occurs! Success is heralded when two of the four detected modes 5-8 detect a single photon (and the other two detectors register no photon). Success can therefore occur ${4 \choose 2}=6$ different ways, the probability of success is 6/32, and the output state on modes 1-4 when it does occur is one of:   
 \begin{eqnarray}\label{eq:multirailencode}
	&& \fket{1,0}\fket{1,0}-\fket{0,1}\fket{0,1}\nonumber \\
	&& \fket{1,0}\fket{0,1}-\fket{0,1}\fket{1,0}\nonumber \\
	&& \fket{1,1}\fket{0,0}-\fket{0,0}\fket{1,1}\nonumber.
\end{eqnarray} 
While these are all maximally entangled, the last possibility would require some shuffling around of modes to get it into the standard qubit dual-rail encoding, and henceforth we will not consider it as a successful generation of a photonic Bell state. As such the success probability reduces to 4/32.

\begin{figure}
\centering
\includegraphics[scale=0.6, clip=true]{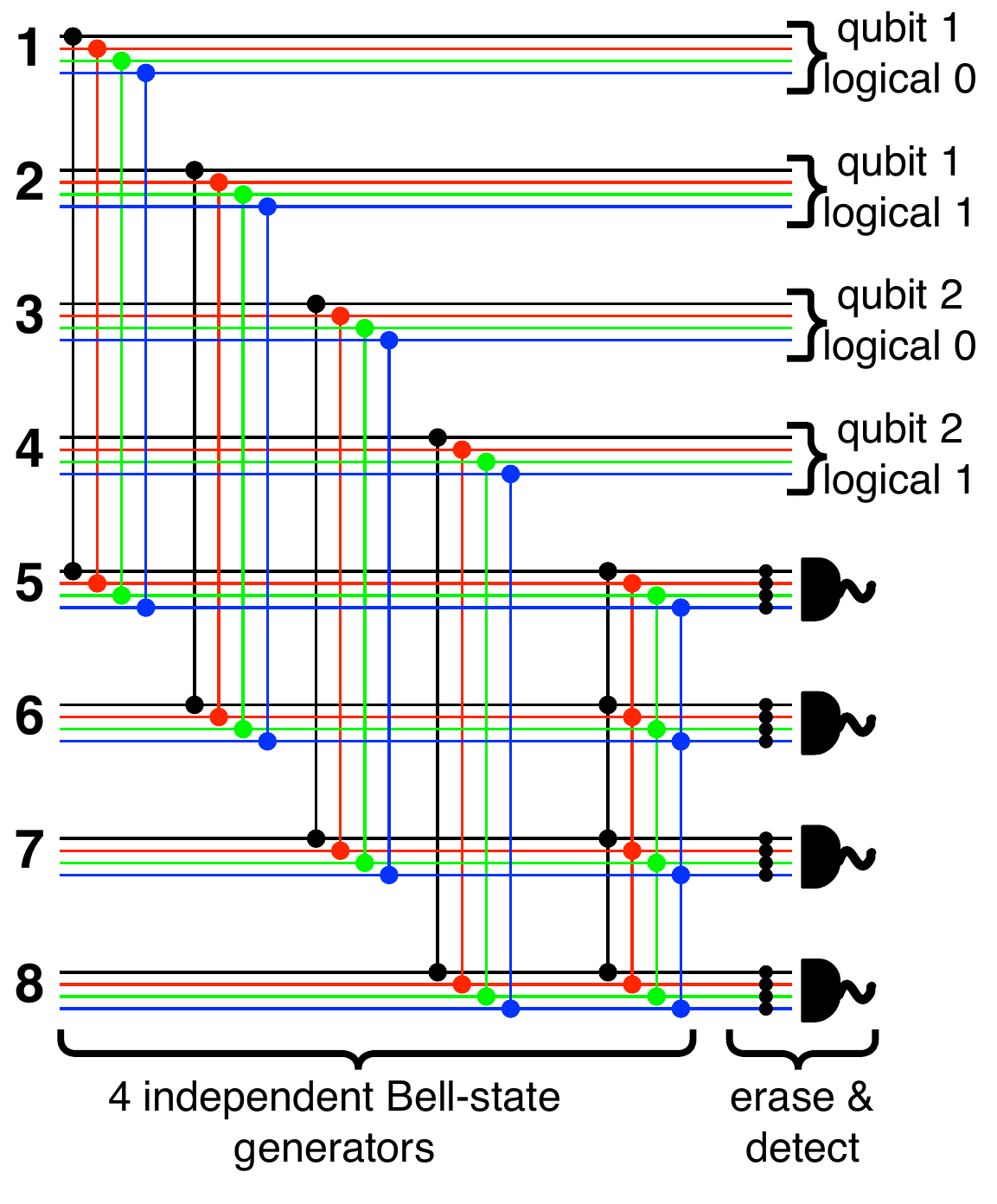}
\caption{Four differently colored copies of the interferometer in Fig.\ref{fig:bellcircuit}(a) can be used to generate multi-rail encoded Bell pairs. Crucially we only need a single photon to enter at any one of the four different colored input modes. That is, rather than requiring a deterministic single photon in mode 1 as per Fig.\ref{fig:bellcircuit}(a), we need only a $[4,1,1]$ source, i.e. a single photon in any one the of black, red, green or blue modes at the inputs labelled 1 (and similarly for inputs 2-4). The interferometers are completely independent until the final beamsplitters labelled ``erasure'', just prior to the detectors. These erase the information as to which of the independent interferometers any detected single photon originates from. The erasure interferometer can have mode transformation $H^{\otimes 2}$ of Fig.\ref{fig:bellcircuit}(b) (although other options exist). Remarkably, we can generalize to $2^k$ copies of the interferometer without loss of success probability!} 
\label{fig:bellcircuitmultirail}
\end{figure}

\begin{figure}
\centering
\includegraphics[scale=0.6, clip=true]{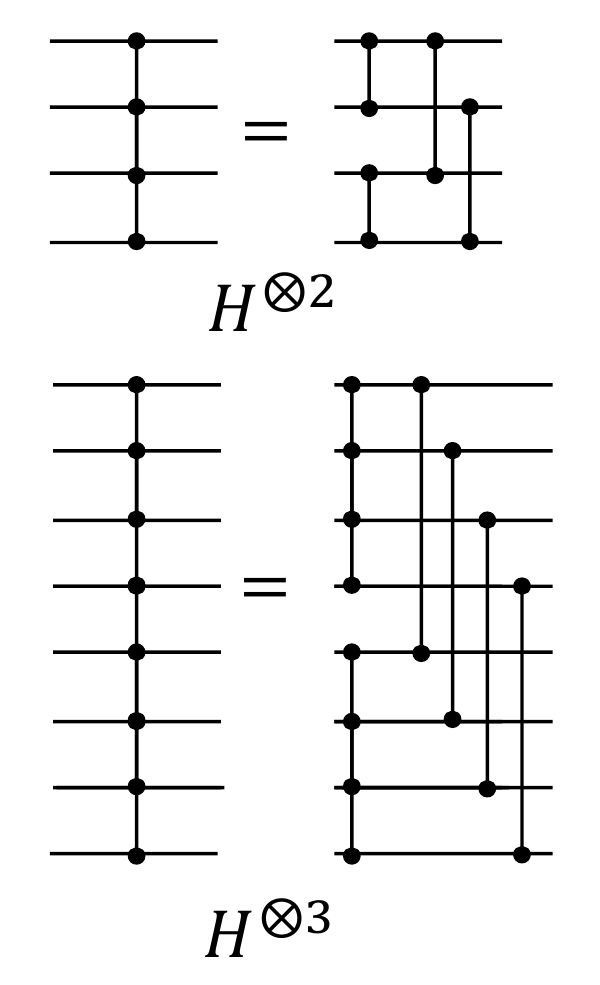}
\caption{Decomposing generalized Hadamard type beamsplitters into regular two-mode 50:50 beamplsitters. The multi-rail architecture requires generalized ``large Hadamard'' beamsplitters. Although a typical interferometer on $n$ modes would take $O(n^2)$ beamsplitters and $O(n^2)$ depth, these can actually be done in logarithmic depth, as the obvious extrapolation of the circuits here make clear. } 
\label{fig:bsrecursion}
\end{figure}

\subsubsection{Production of multi-rail encoded Bell pairs using stochastic sources} 

In this section we overview a method to generate multi-rail Bell states using stochastic sources. The scheme has almost the same success probability as for the dual rail case. Quite remarkably, the success probability does \emph{not} depend on the size (i.e. total number of modes) of the stochastic source - the non-determinism of the source only requires construction of a larger interferometer.

An example of such multi-rail Bell state generation using four stochastic sources is given in Fig.~\ref{fig:bellcircuitmultirail}. A $[4,1;1]$-source at each of the inputs labelled 1-4, will lead to successful generation of a multi-rail encoded Bell pair across multi-rails $1,2$ for the first qubit, and $3,4$ for the second\footnote{In fact the Bell generation circuit doesn't care if we input the single photon into mode 1 or into mode 5 - and similarly for the 2,6 or 3,7 or 4,8 pairs. As such we could use a $[8,1;1]$ source across all the modes in each pair, something which definitely improves resource counting  but it complicates the exposition.}. The exact success probability depends on the particular modes randomly occupied by each of the four input photons, but we find it is never less than half of that obtained for the original scheme of Fig.~\ref{fig:bellcircuit} using deterministic photons, i.e it is at least $2/32$. This remains true even if we construct the obvious generalizations to $8,16,32\ldots$ copies of the basic interferometer in order to deal with larger stochastic source inputs - that is, there is no further decrease in success probability no matter how many copies we use!\footnote{It was this surprising observation - which ultimately arises because we are using interferometers restricted to only $\pm 1$ phases that prevents the interference ``going anywhere else'' - which predicated the discovery of these architectures.}      

Particularly intriguing are the cases where the four input photons happen to each enter
a completely different interferometer. Success is heralded by one and only one photon at the detector(s), meaning that parts of the total quantum state that lead to success must have had only one photon entering the erasure beamsplitter just prior to the detector. Thus it becomes clear that no multi-photon interference need occur as part of the successful Bell state generation. By increasing the number of rails (copies of the initial interferometer) we could ensure vanishing amplitudes for multi-photon interference in both the success and failure parts of the wavefunction. All that is required is interference/erasure of ``Feynman paths'', not the bosonic nature of the photons, and therefore there are numerous other physical systems in which one can imagine implementing these basic procedures.

\begin{figure}
\centering
\includegraphics[scale=0.5, clip=true]{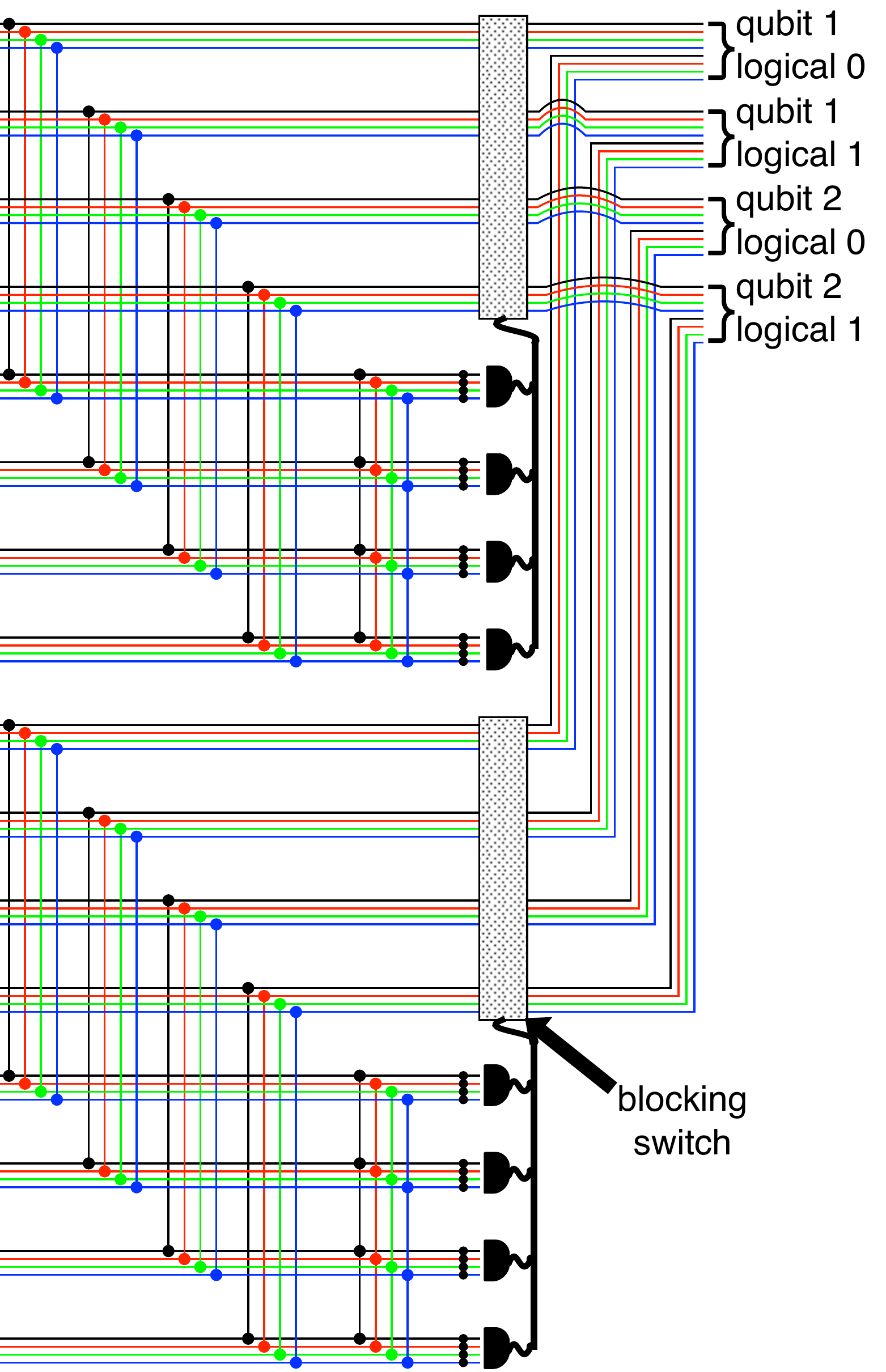}
\caption{By increasing the number of rails over which the multi-rail qubit is encoded we can `parallelize' and effectively `passively multiplex' the entangled state production from many independent Bell state generators up to near unit efficiency without having to do any coherent switching at all. We do need to use a blocking (absorbing/filtering) incoherent switch in order to ensure extraneous photons from failed generators do not contaminate the multi-rail output. Here a purely-spatial passive multiplexing of two generators is depicted. Either (or both) of the Bell state generators could produce a multi-rail Bell pair with each qubit defined in a four spatial mode encoding, after the passive multiplexing the output qubits are defined across an eight mode encoding.}
\label{fig:2bells8multirail} 
\end{figure}

\subsection{Avoiding switches via ``passive multiplexing''}

As the entangled state generation just presented is probabilistic it is tempting to think we must now resort to standard multiplexing: using a coherent multimode switch to shuffle our spatially multi-rail encoded qubits to be dual rail encoded, and furthermore making many copies of the Bell state generator and using a switch to select out one which has succeeded. 

Later we will show, however, that the remaining primitives of photonic quantum computing (fusion gates and creation of arbitrarily large cluster states) can all be performed via passive interferometers on multi-rail encoded qubits with the same probability they can be performed on dual-rail qubits. That is, the size of the multi-rail encoding does not affect the success probabilities, it changes only the size of the photonic circuits. Since we therefore need not (in principle!) care how many modes the multirail encoding is distributed over, we can simply build many such Bell state generators and trivially combine the outputs to make a Bell state with a larger multi-rail encoding. This is depicted in the simplest case of two spatially parallel Bell state generators in Fig.~\ref{fig:2bells8multirail}.

We can further imagine repeatedly firing many multi-rail Bell state generators and interpreting the output as a spatio-temporal stochastic source of multi-rail encoded Bell pairs. The efficiency of each is low (around 2/32), but 
(similar to the situation with single photons) there are various options to combine many such low-efficiency devices to create a near-deterministic Bell state source. This new source will generically be of a mixed spatio-temporal character, although a purely spatially encoded source could be built by generalizing Fig.~\ref{fig:2bells8multirail}. Note that failure outcomes for generating the Bell states leave undesired photons in output modes that would need to be filtered with a blocking type of switch, similar to the case of building a stochastic single photon source discussed at the start.

\subsection{Killing Time}\label{subsec:killingtime}

A crucial element of our ability to use a (purely spatial) stochastic single photon source to generate Bell states is the existence of an ``erasing'' interferometer for spatial modes, namely the one described by the mode transformation $H^{\otimes k}$. If we want to make use of mixed spatio-temporal sources we need to consider the extent to which similar procedures exist for erasure of temporal mode information. Such erasure will also be important for doing fusion gates with multi-rail qubits of a mixed spatiotemporal character.

In the Bell state generation of Fig.~\ref{fig:bellcircuitmultirail} we begin with four separate spatial interferometers, copies of the regular Bell state generator of Fig.~\ref{fig:bellcircuit}. Consider a different scenario, where four stochastic sources of $[1,M,1]$ type (for some $M\gg 1$) photons are used at the four inputs of Fig.~\ref{fig:bellcircuit}. Then we effectively are populating $M$ independent copies of the Bell state generator. To mimic the structure of Fig.~\ref{fig:bellcircuitmultirail}  - which erased spatial information via an $H^{\otimes k}$ interferometer - we need an analogous method of erasing temporal information.   

Typically manipulating quantum information using time-bin encodings necessitates the use of ``active'' (dynamical) elements, and our overarching goal here is to avoid such components. As we now show, it is possible to suitably \emph{erase} temporal information using passive generalizations of Franson\cite{FransonInterferometer} interferometers, although there is a finite chance of (heralded) failure. 

\begin{figure}
\centering
\includegraphics[scale=0.6, clip=true]{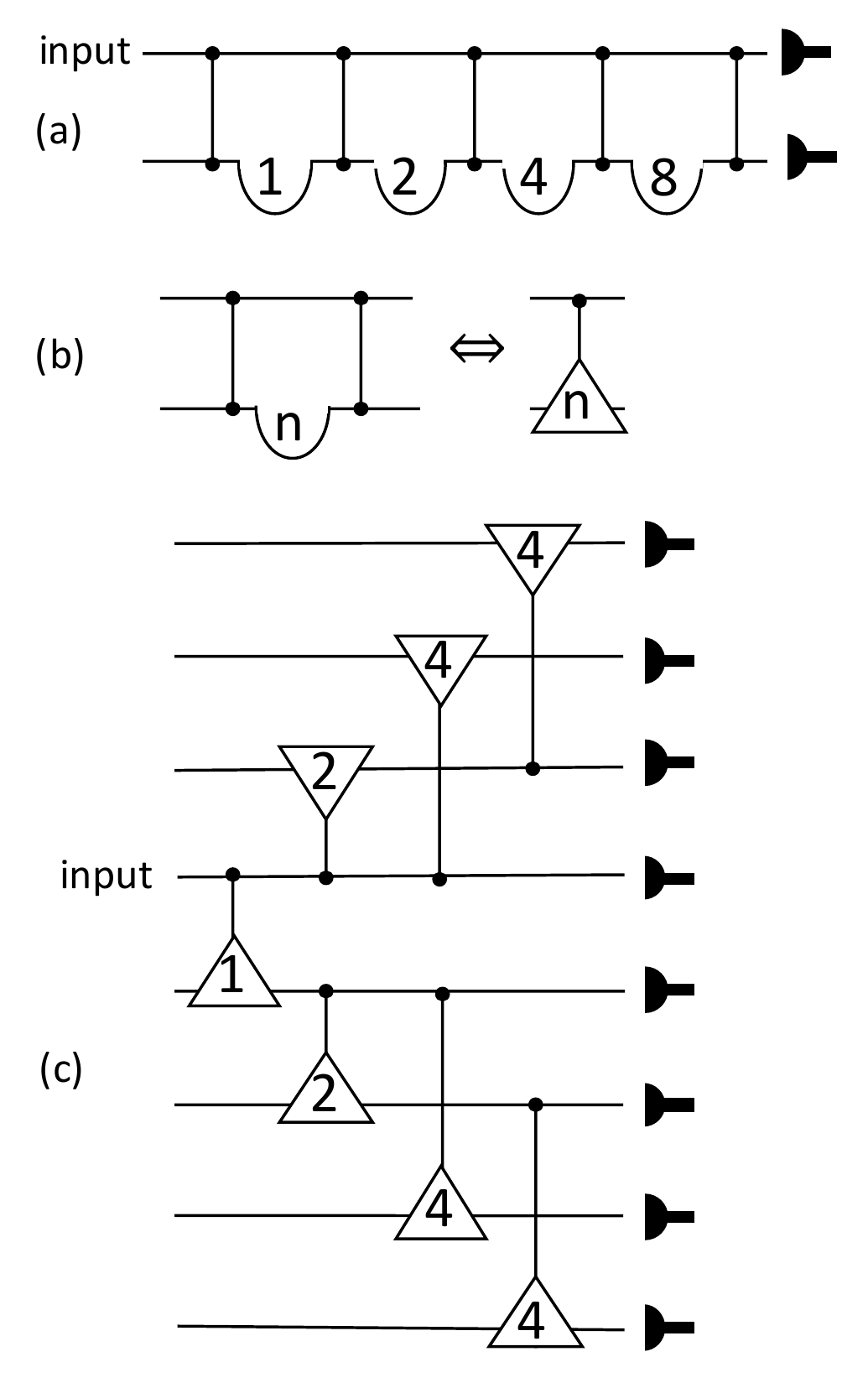}
\caption{Examples of passive interferometers that allow for erasure of a single photon's timing information with arbitrary (heralded) probability of success (when extrapolated in the obvious pattern). (a) A photon in a single spatial mode is coupled to one other empty ancilla spatial mode via a series of beamplsitters. One spatial mode is delayed, the integer indicates the number of time bins each delay takes. The output state is a uniform superposition over 16 time bins and the two spatial modes. (b) A simplified notation for two beamsplitters and a delay of $n$ timesbins. (c) Another interferometer that erases timing information, but spreads the input over increasing numbers of spatial modes.} 
\label{fig:temporalerasure}
\end{figure}

To understand the general strategy, consider a single photon in one spatial mode, but potentially in any one of a finite set of time bins $t_i$. Imagine we can uniformly `spread' the photon across many spatial and temporal modes:  
\begin{eqnarray}
\fket{1_{\vec{k}_1,t_i}}\,\,\,\longrightarrow \sum_{a,b} C_{a,b}(i) \fket{1_{\vec{k}_a,t_b}}, 	
\end{eqnarray}
where by `many' we mean the discrete indices $a,b$ have combined range much greater than the range of $i$.
As long as the coefficients $C_{a,b}(i)$ are uniform in magnitude, and as long as $C_{a,b}(i)\neq 0$ for all initial time bins $t_i$ over a large range of $a,b$, then when we detect the photon we will have erased our knowledge of which time bin $t_i$ it originated in. Note that the phases of the $C_{a,b}(i)$ are important to know since our photon is typically entangled with others.\footnote{This is why erasing time bin information using jittery detectors etc is not sufficient! We require \emph{coherent erasure}.}  Some examples of interferometers which achieve this are shown in Fig.~\ref{fig:temporalerasure}. For these examples there will be some values of the detected spatial mode/time bin (i.e values of $\vec{k}_a,t_b$) which can only arise from a single initial time bin, or more generally from less than the full set of initial time bins $t_i$ of interest, and such detections amount to a failure of the temporal erasure.

The upshot of all this is that we can readily combine erasure of both spatial and temporal information. This may or may not be useful for the generation of the initial entangled states. However, as mentioned above, with repeated use of hardware the entangled photons we produce can often be interpreted as be multi-rail encoded, where the modes comprising the rails vary both spatially and temporally. 

We now have the tools in hand to consider using multi-rail entangled states directly for quantum computing, that is, without using coherent switches to ``compress down'' the photons into a smaller number of spatio-temporal modes as per the standard multiplexing ideas.

\subsection{Multi-Rail Fusion Gates}

\begin{figure}
\centering
\includegraphics[scale=0.6, clip=true]{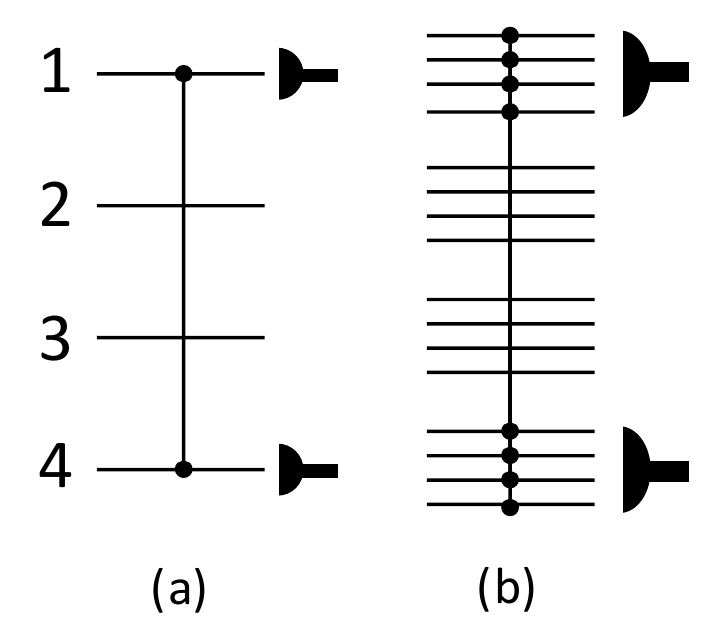}
\caption{(a) A regular Type-I fusion gate on a dual rail encoding is simply a beamsplitter between one rail of each of the two dual rail qubits encoded in modes 1,2 and 3,4 respectively. When the gate succeeds the output qubit is encoded across modes 3,2. (b) On  spatially multi-rail encoded qubits a general Type-I gate can be achieved using the generalized ``large Hadamard'' beamsplitters introduced earlier, which completely erase spatial mode information. If the photons are also spread across different time bins then these can be erased independently, by placing interferometers such as those in Fig.~\ref{fig:temporalerasure} just prior to the detectors.} 
\label{fig:multirailfusion}
\end{figure}

\subsubsection{Type-I fusion}

The question to which we now turn is whether it is possible to use a multi-rail encoded entangled state as part of a full quantum computation. In the standard architecture we build larger entangled (cluster) states by fusing \cite{Browne2005} together smaller ones. 

The simplest variant of a fusion is performed by the Type-I gate. On dual rail photonic qubits it amounts to simply a beasmplitter between the first mode of qubit 1 and the last mode of qubit 2. This generates the evolution:
\begin{eqnarray}  
\ket{0}\ket{0}=\fket{10}\fket{10}&\rightarrow & \fket{10}\fket{10}+\fket{00}\fket{11} \\
\ket{0}\ket{1}=\fket{10}\fket{01}&\rightarrow & \fket{20}\fket{00}-\fket{00}\fket{02} \\
\ket{1}\ket{0}=\fket{01}\fket{10}&\rightarrow & \fket{01}\fket{10} \\
\ket{1}\ket{1}=\fket{01}\fket{01}&\rightarrow & \fket{11}\fket{00}-\fket{01}\fket{01}.	
\end{eqnarray}
From these expressions we see that the only way the detectors in modes 1 and 4 can both detect one and only one photon is if the initial qubits were in the same logical state. Moreover, when this occurs we have no information about whether they were both $\ket{0}\ket{0}$ or whether both $\ket{1}\ket{1}$ - the beamsplitter renders both options equally likely - but the middle two modes 2,3 are now left in a new dual-rail encoded qubit state, that can be interpreted as logical $\ket{0}$ if the original two qubits were $\ket{0}\ket{0}$ and logical $\ket{1}$ if the original two qubits were $\ket{1}\ket{1}$. When the initial two qubits are in different logical states, we can tell from the detection pattern in modes 1 and 4 (either no photons at all, or two photons detected) which of the two cases pertains. 

When the action of the fusion gate is described in words as in the preceding paragraph, it becomes essentially obvious how a multi-rail encoded version of the gate can be implemented - the paragraph could be re-read replacing ``mode 1'' by ``multi-rail mode 1'' and so on. All that is required is a generalization of the beamsplitter between modes 1,4. That is, we want a device that, in the case where there is only a single photon in one or other of these two modes, can detect that fact without revealing which particular mode it originated from. But this is exactly what we can achieve for purely-spatial multirail encodings using the $H^{\otimes k}$ interferometers, for temporal encodings using the techniques in Section \ref{subsec:killingtime} above, and for mixed encodings by combining the two.

It should be emphasized again that multi-photon interference is not strictly necessary, which is essentially why we can do spatial erasure and then temporal erasure - we are only ever selecting as success the occasions when a single photon went through the erasing device(s). 

\begin{figure}
\centering
\includegraphics[scale=0.6, clip=true]{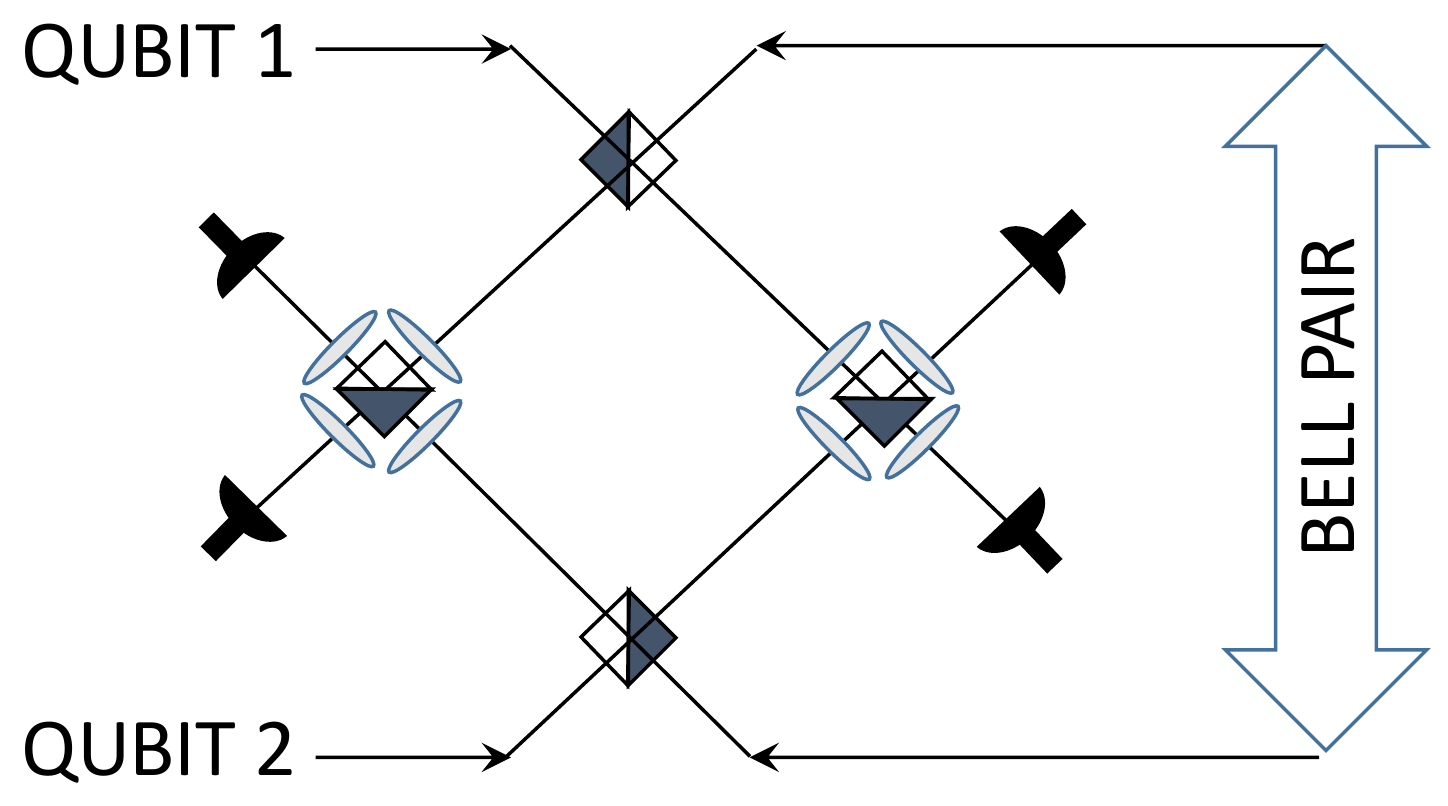}
\caption{Boosted Type-II gate, as in the original \cite{Grice2011} here we revert to mixed spatial and polarization encoding which makes the following explanation of the correlations observed a little simpler: A Bell pair in the state $\ket{HH}+\ket{VV}$ is injected as an ancilla, along with the two qubits. If the two qubits have opposite polarizations then one photon goes to the left and one to the right, while both photons from the ancillary Bell pair go the same way. Thus if we erase the information as to the origin of the photons (which is what the arrangement of a PBS surrounded by $45^\circ$ degree polarization rotators achieves on both the left and the right), and count 3 on one side and 1 on the other, we know the qubits had opposite polarization (but not what it was). If the qubits have the same polarization they both go left or both go right. If it so happens that the anciallary photons go the same way then the gate fails - we know for sure what the qubits polarization was. However if the ancillary photons go to the opposte side to the qubits, and if the distinguishing features of the photons are removed the gate succeeds because we count 2 photons on the left and 2 on the right, which heralds only that the photons were the same polarization.} 
\label{fig:boostedII}
\end{figure}

\subsubsection{Beyond Type-I}

Type-I fusion works with probability 1/2; if we need to erase temporal as well as spatial information it will work with slightly lower probability (dependent on how much we `stretch' the photons temporally to erase them). We can use Type I fusion to create (passively multiplexed) versions of multi-rail GHZ states. 

It is often convenient to use `boosted' Type-II fusion gates\cite{Gimeno-Segovia2015} Boosted fusion \cite{Grice2011} involves an ancillary Bell pair of photons. the four photons in 8 modes evolve through the interferometer into a quantum state containing thousands of terms, many of which have multi-photon interference as exhibited by terms involving Fock states $\fket{2},\fket{3},\fket{4}$, and unlike the cases above the ``success'' outcomes often involve detection of such multiphoton events. As such it would seem unreasonable to presume that it can be trivially performed in the multi-rail way. But for the variant which involves failure outcomes that project into the computational basis the way it works can be ``talked through'' almost as simply as the case for Type-I fusion above. This is done in Fig.~\ref{fig:boostedII}. One we appreciate that the gate relies only on eliminating the information about origin of the photons that enter the detector - the polarizing beamsplitter surrounded by rotators is simply the generalized beamsplitter $H^{\otimes k}$ gate- it becomes clear that the gate works just as well on a multi-rail qubit (though potentially necessitating a much larger version of such a Hadamard interferometer!)

\subsection{Creating multi-rail cluster states despite no single qubit Hadamard gates}

For simplicity in this section we focus on purely spatially multi-rail states. 

Because of the impossibility of doing arbitrary single qubit gates on a generic multi-rail qubit, and because we have only shown how to create Bell/GHZ states and fusion gates (that fuse in the computational basis), it is still not given that we can actually create suitable multi-rail versions of a cluster state.  

To spell the issue out in more detail: Consider we do a beamsplitter to perform the supposed Hadamard gate on an $m=2$ rail  Bell state, where the rails have ``misaligned'' photons:
 \begin{eqnarray}
	&& \ket{0}\fket{10}\fket{00}+\ket{1}\fket{00}\fket{01}\nonumber \\
	&\rightarrow & \ket{0}(\fket{10}\fket{00}+\fket{00}\fket{10})\nonumber \\ && +\ket{1}(\fket{01}\fket{00}-\fket{00}\fket{01})\nonumber\\
&\neq & \ket{0}\ket{+}+\ket{1}\ket{-})\nonumber
\end{eqnarray}
We see that unless we allowed for different qubit encodings for different parts of a superposition (we do not - such really would stretch the notion of a qubit!) naive application of a Hadamard cannot create the desired 2-qubit cluster state. Not unexpectedly, the multi-rail nature of the state has interfered with interference.   
 
One simple solution to this is to modify the generator which produced Bell states from 4 single photons to one which produces 2-qubit cluster state directly. This can, in fact, be done by replacing the $H^{\otimes 2}$ interferometer in modes 5-8 of Fig.~\ref{fig:bellcircuit}(a) with an interferometer described by the following unitary matrix:
  \[
\frac{1}{2} 
\begin{bmatrix}
1 &i &-1 &i \\ 1 &i &1 &-i \\-1 &i &i &-1 \\1 &-i &i &-1
\end{bmatrix}.
\]
For the regular case where we have four single photons input to modes 1-4 this will produce dual-rail encoded 2-qubit cluster state with probability $2/32$. Note that we have not done a single quit gate on the output modes, rather we are doing judiciously chosen quantum steering to collapse the outputs to the desired state.

When we incorporate this modified generator into a multirail architecture with stochastic sources, following (the arbitrarily large generalization of) Fig.~\ref{fig:bellcircuitmultirail}, depending on the exact modes the photons are input the success probability may fall to $1/32$, but it is never lower than this. 

Once we have pieces of linear cluster state we can proceed to use (boosted) fusion to percolate a lattice - the pieces of the cluster state now have the Hadamard transformation `built in', so the fusion will work with probability $3/4$ as described previously. From there the creation of arbitrarily large cluster states is immediate.

\subsection{Adaptive single qubit measurements}
 
Performing single qubit unitary rotations about the $Z$-axis is simple on multi-rail qubits, it involves implementing phase shifters across all the relevant modes defining the qubit.  As such computation using a multi-rail cluster state reduces to whether we have the ability to make both $X$ and $Z$ measurements. Note that The $Z$ measurements are obviously unaffected by the multi-rail nature of the state. 

The $X$ measurements can be performed using the generalized Hadamard beamsplitters $H^{\otimes k}$. To see this consider the example:
\[
\ket{A_0}\fket{10}\fket{00}+\ket{A_1}\fket{00}\fket{01},
\]
where the 4 modes encode an $m=2$ multi-rail qubit.
 An $X$ measurement on the qubit that has been singled out should leave the remainder of the cluster state in $\ket{A_0}\pm\ket{A_1}$. After the $H^{\otimes 2}$ beamsplitter we have 
 \begin{eqnarray*}
 && \ket{A_0}(\fket{10}\fket{00}+\fket{01}\fket{00}+\fket{00}\fket{10}+\fket{00}\fket{01}) \\
  & +&\ket{A_1}(\fket{10}\fket{00}-\fket{01}\fket{00}-\fket{00}\fket{10}+\fket{00}\fket{01})
 \end{eqnarray*}
 and so regardless of which mode we detect the photon in, we obtain one of the desired collapsed states.
  
We need, however, to be able to \emph{adaptively} choose whether we perform an $X$ versus a $Z$ measurement. Surely this requires a coherent switch? In principle it does not: we can set up a static interferometer as if we are doing an $X$ measurement, let the photons through if that is what we want to do, but absorb (with a blocking switch) the set of rails encoding a $\ket{0}$ at the input if we prefer to do a $Z$ measurement. In the latter case then absence of a detected photon would correspond to the $\ket{0}$ outcome while detection would correspond to a $\ket{1}$.\footnote{It should be emphasized that while this allows us to in principle only use a blocking switch throughout, in practice it completely negates one of the best features of dual/multirail encodings, namely that photon loss (the dominant error mechanism) becomes heralded. Such loss would now lead to Pauli error!}
  
Once we allow for mixed spatio-temporal encodings things become a bit less efficient, but the basic principles remain the same. Although it should be said that if the photons were in a \emph{purely} temporal multirail encoding (e.g. a 3-photon GHZ with all three photons in the same spatial mode) things become very messy. However our Bell state generation does not produce such entangled states - they always come out across multiple spatial as well as temporal modes. Moreover, given our ability to manipulate spatial degrees of freedom better than temporal ones, we should presumably only consider the case that at least two spatial modes are involved in encoding a multi-rail qubit.


%
%


\section{Paper II: A modular, networked quantum computer based upon qubits fully delocalised across the network nodes.}

In a standard networked architecture for a quantum computer qubits are localized at nodes of the network, and can interact readily with other qubits at the same node. Interactions between qubits at different nodes are then either mediated by exchange of quantum information (normally photons) or via teleportation (perhaps with ancilla entangled qubits distributed continuously from a central resource). Typically all nodes send measurement results and receive instructions by exchanging classical information with a central controller (classical computer).

\begin{figure}
\includegraphics[scale=0.8, clip=true,bb= 2.9cm 15.7cm 15cm 24.5cm]{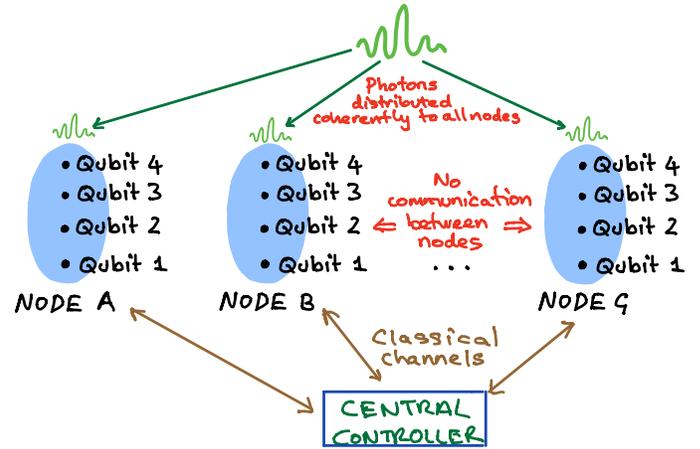}
\caption{Schematic of a delocalized networked architecture for a quantum computer.} 
\label{fig:DNAfig}
\end{figure}

The delocalized network architecture (DNA) we outline in this paper uses delocalized photonic qubits, such that every qubit has some ``piece'' (amplitude for being located) at every node (Fig.~\ref{fig:DNAfig}). Interactions between qubits are still performed locally, but each interaction happens at every node (and all nodes are structurally identical). When measurements are performed the qubit (presumed destroyed) is, of course, only detected at one of the nodes. Classical information is sent to a central controller (by whichever node did detect the qubit) regarding the measurement outcome, and the central controller transmits classical instructions back to all nodes regarding operations to be performed on the remaining qubits. No other quantum information exchange between nodes is required. 

The only quantum information exchange is the distribution of single photons to the nodes, where each single photon is in a superposition of being sent to every node. This step is the entanglement distribution to the network. 

A natural objection might be: ``Standard photonic quantum computing depends only on having ``identical'' photons, by which we actually mean we only need photons that can be decomposed
\[
\left(\sum_\alpha c_\alpha a^\dagger_\alpha \fket{0}\right)\, \otimes \,\,\left(\sum_\beta c_\beta a^\dagger_\beta \fket{0}\right) \,\, \otimes \ldots
\]
where the expansion coefficients $c_\alpha, c_\beta$ are completely arbitrary, though they need to be the same (even if not known!). Therefore if we let the modes $\alpha,\beta$ be discretized and spatial, we can obviously do a distributed computation of the form described above.''

Note the single photon states used in the objection above all have overlap 1. What we will see in the DNA is that when $|c_\alpha|^2=|c_\beta|^2=\ldots$ (ie uniform magnitude photonic states) we can still do the distributed, delocalized computation even if the phases of the $c_\alpha, c_\beta$ are such that the states in question  are completely orthogonal!  

Interestingly, phase stability between the network nodes is not required, although phase stability between all modes held by any given  node must be maintained. Moreover, each of the nodes can itself be split into sub-nodes to which quantum information is sent, and phase stability need only be maintained between the modes entering a given sub-node (a feature which repeats itself - the depth of sub-nodes is determined by the number of probabilistic elements used within any particular choice of photonic architecture). 
      
 As in Paper I, a surprising feature is that in principle this architecture only requires a `blocking' (absorbing/filtering/incoherent) switches. In fact one variant of the DNA shows that (in principle) all but one of the blocking switches do not even need to act on the single modes in which quantum information is stored - they can act only on the classical pump fields which drive the single photon sources! While this extreme variant requires very long (fixed) delay lines, it shows that it is in principle possible to do photonic quantum computing in an architecture where every photon passes through a constant-depth, passive interferometer followed by just one switch just prior to  being detected.

\subsection{High Level Summary}\label{sec:highleveloverview}

To set some terminology we first consider a regular non-distributed photonic architecture. We call a photonic quantum computer which can take in deterministic single photons and perform a quantum computation a level 0 quantum computer. We call a photonic quantum computer which can take in deterministic photons entangled over a small number of qubits (Bell pairs, GHZ states, something else depending on the specific proposal) and perform a quantum computation a level 1 quantum computer. We call a photonic quantum computer which can take in deterministic photons entangled over a large number of qubits (percolated cluster state lattice, something else depending on the specific proposal) and perform a quantum computation a level 2 quantum computer. For certain variants there may be more intermediate levels - e.g. Bell pairs being fused into small GHZ states would increase the depth, but for simplicity we stick to levels 0,1,2 computers. 

Regular photonic architectures use multiplexing (performing an operation many times in parallel and careful selection and switching out of successful outputs) to compress out the inherent randomness of photonic gates/processes. Multiplexing allows one to nest a level 1 computer into the level 0 computer, and a level 2 computer into the level 1 computer. Multiplexing is also an option to create the initial deterministic single photons (although of course many other options exist for deterministic photon production). 

At a practical level multiplexing photons requires the ability to switch coherently - at a minimum to be able to actively select $1$ from $m$ modes and coherently (i.e. preserving the quantum state) transfer the light within that mode  to some other part of the computer. It is a dynamic process which `removes the randomness', as it were, by effectively erasing the information as to which particular probabilistic element it was that succeeded. As such, from now on we will refer to such a process as \emph{dynamic multiplexing}.

\begin{figure}
\includegraphics[scale=0.6, clip=true,bb= 3.5cm 10cm 14cm 25.5cm]{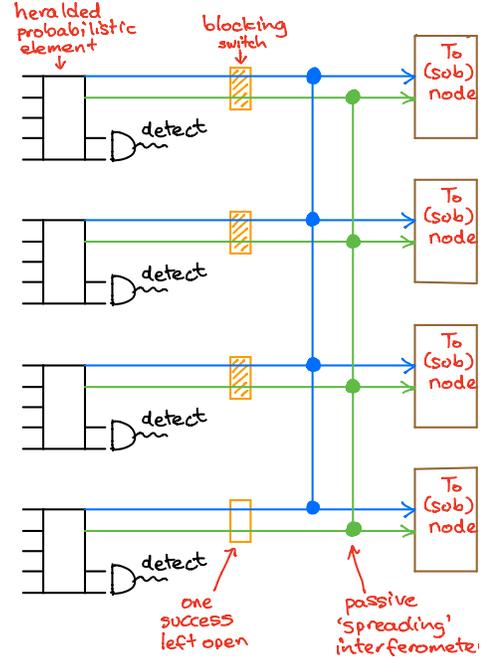}
\caption{The generic ``passive multiplexing'' procedure replacing dynamic multiplexing in the DNA. Horizontal lines denote modes \emph{not} qubits. The output modes from all but one successful process are blocked. After the blocking switches the passive interferometers which spread the output do so uniformly. That is, the unitary matrix describing such interferometers has complex entries of equal magnitude, such as a discrete fourier transform or $H\otimes H$, with $H$ the $2\times 2$ Hadamard matrix.} 
\label{fig:basicprocedurefig}
\end{figure}

 The DNA uses a different approach to dealing with a probabilistic processes. The generic procedure is depicted in Fig.~\ref{fig:basicprocedurefig}. Multiple copies of the process are performed, such that there is high probability that at least one of them succeeds. All outputs, except those of one of the successful processes, are blocked. We then uniformly ``spread'' across all of the potential outputs of all of the probabilistic elements as depicted. The spreading is performed by a passive interferometer, and it is this process which `removes the randomness', as it were, effectively erasing (from the state amplitudes) the information as to which particular probabilistic element it was that succeeded. Of course information as to which element succeeded remains in the (known) relative phases between modes of the output state. If we restricted to using a power of two copies of the probabilistic process then the passive interferometers could be taken to be specified by the unitary matrix $H^{\otimes k}$ where $H$ is the 2d Hadamard matrix, and relative phases are all $0$ or $\pi$. In general other uniformly spreading interferometers such as the discrete fourier transform can be used. From now on we will refer to `removing the randomness', as it were, by the procedure of Fig.~\ref{fig:basicprocedurefig} as \emph{passive-multiplexing}.
  
The outputs are passed to sub-nodes (higher level quantum computers) which are built to treat the incoming modes \emph{as if they deterministically contained the desired state}. As with the nodes themselves, the subsequent sub-nodes do not exchange any quantum information and so do not need to be kept phase stable with respect to each other, a feature of practical importance.  


 Once this generic procedure is implemented within sub-nodes of the DNA it can be incredibly confusing to keep track of levels and modes and nodes and so on. For this reason we turn to a highly idealized example that captures the basic quantum and classical information flow within the DNA.

 \subsection{Medium Level Summary}\label{sec:mediumleveloverview}

To illustrate the rather complicated manner in which the photons are delocalized through several stages, it is helpful to imagine we have a standard photonic architecture with the following (very highly idealized!) properties:
 
 (i) We can produce single photons from stochastic sources with a high enough probability that if we have $7$ single photon sources it is almost certain that at least one produces a photon.   
 
 (ii) Bell pairs can be produced from $4$ single photons with a high enough probability that we need only make $3$ attempts to produce a Bell pair and then we will have one with almost certainty.
 
 (iii) We are able to fuse together Bell pairs with a high enough probability that we need only 5 Bell pairs and we will almost certainly produce a large enough entangled state that we can solve the problem we wish to tackle on our quantum computer(!). Solving that problem will necessitate being able to choose between $X$ and $Z$ single qubit measurements. 
 
We now consider how the DNA would look built using the same processes, but replacing dynamic with passive multiplexing.
 
 \subsubsection{Step 1 - Distribution of single photons}
 
 The DNA will have $7$ nodes, labelled $A,B,C,\ldots,G$, each of which is a self-contained level 0 computer, i.e. they treat their incoming single modes \emph{as if} they contained deterministic single photons. 
  
\begin{figure}[t]
\includegraphics[scale=0.7, clip=true, bb= 3.7cm 10.5cm 15.2cm 25.1cm]{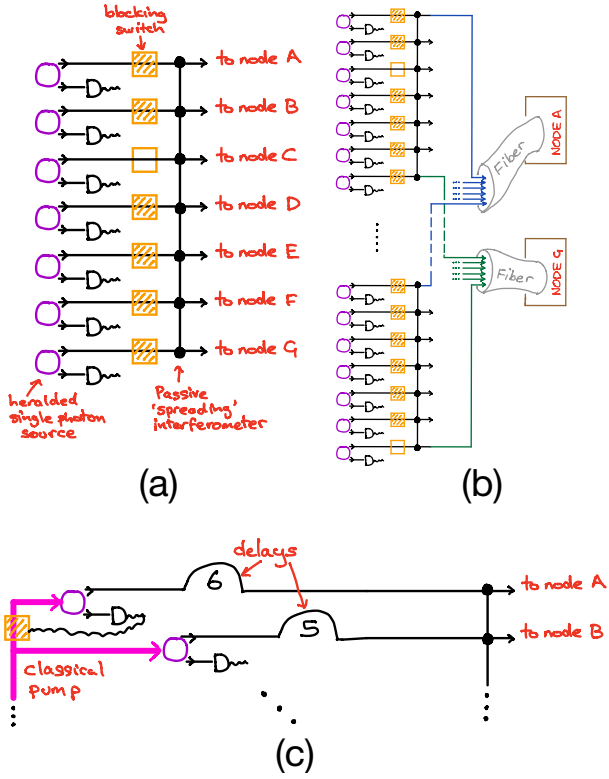}
\caption{ (a) A `deterministic' single photon source built from (heralded) stochastic sources. Based on the heralding pattern the output from one of the successful sources (in this example the third source) is passed through onto a passive `spreading' interferometer. (b) The output from as many single photon sources as required in the computation (in this example 420) are distributed to the network nodes. All modes going to a given network need to be kept phase stable with respect to each other, but stability between the nodes is not necessary. This raises the possibility of well-separated nodes, receiving modes distributed via optical fiber(s), as long as the noise processes within are collective (ie act identically on the propagating modes).  (c) An alternative to blocking the single photon output modes of a passively multiplexed source. With suitable use of fixed delays (of up to $6$ clock cycles for this example) we could attempt to produce a photon at the first source and if it succeeded turn off the classical \emph{pump} laser to all the subsequent sources. If it failed we would then attempt to fire the second source and so on. This has the advantage of only needing a blocking switch on the classical pump and not the quantum system that is eventually part of our computer.}
\label{fig:singlephotonsource}
\end{figure}

 We begin with  $420=7\times 4 \times 3 \times 5$ single photon sources. Our ``effectively deterministic'' single photon source is built by taking the stochastic sources in groups of $7$ at a time, and passively multiplexing them as depicted in Fig.~\ref{fig:singlephotonsource}(a).
 
The central controller needs to keep a record of which specific stochastic sources fired, since there are important relative phases between the output modes of each deterministic source. However, it is fine for all of the modes which are passed to a given node to receive the same random phase (such as might occur if taken to their destination node along the same multi-core optical fiber, Fig.~\ref{fig:singlephotonsource}(b)), uncorrelated with the random phases on groups of modes being passed to other nodes. This is explained in more detail later.

 \begin{figure}
\includegraphics[scale=0.6, clip=true,bb=4.5cm 4.2cm 17cm 25.3cm]{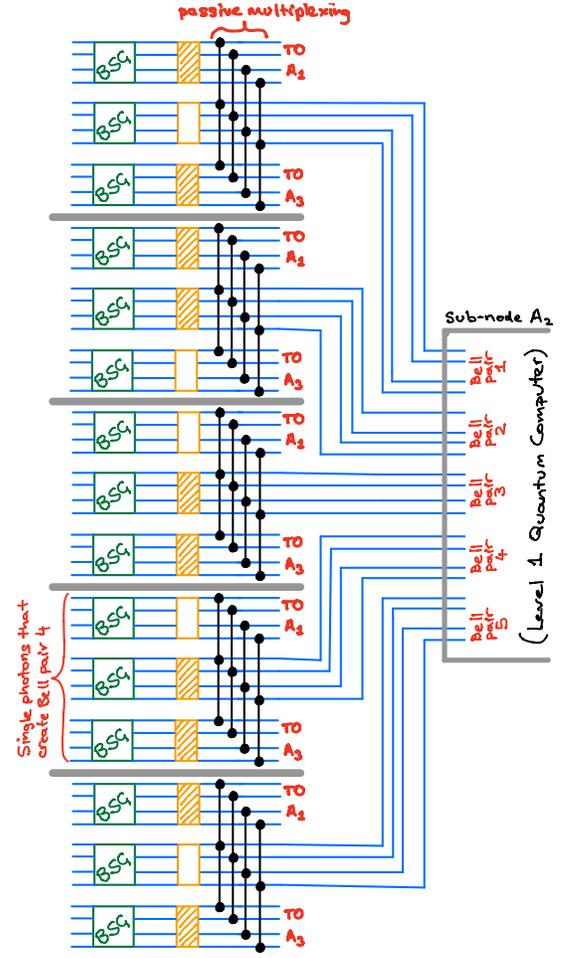}
\caption{Bell state generation and passive multiplexing within node A, which is a Level 0 quantum computer - that is, it treats incoming single photon modes \emph{as if} they held a deterministic single photon.}
\label{fig:nodeAfigure}
\end{figure}

\subsubsection{Step 2 - Creation of Bell pairs}
 
For concreteness we describe the level 0 computer of node $A$, however all other nodes are built and function identically. Node $A$ has $4 \times 3 \times 5$ incoming modes, which it treats as if they contain a deterministic single photon (although if actually measured at this stage there is only a $1/7$ probability of finding any given mode occupied because any given photon has equal likelihood of ultimately being detected at any of the nodes).  

The incoming single photon modes are gathered into groups of 4 and each group sent through a probabilistic Bell-state generator (BSG), see Fig.~\ref{fig:nodeAfigure}. In our toy example, $3$ such BSG's are required for almost certain generation of a Bell pair. As such, Node $A$ takes groups of $3$ BSG's and uses passive multiplexing to produce `deterministic' Bell pairs. The output from one successful BSG is not blocked, and thus enters the spreading interferometer. As depicted in Fig.~\ref{fig:nodeAfigure} the outputs are then sent to sub-nodes of the DNA, labelled $A_1, A_2, A_3$. These sub-nodes are level 1 quantum computers - that is, they are built to treat each incoming group of 4 modes as if they contained a deterministic Bell pair. As such they would implement some kind of fusion gates on these modes. Each of the sub-nodes operates independently (no quantum information is exchanged with any other sub-nodes) and so it would not matter if all the modes entering a given sub-node received the same unknown phase (e.g. during transmission to the sub-node from the higher node).  

 At this stage one may wonder how it can be that BSG's with potentially no incoming photons can herald successful generation of a Bell pair? This is where the Central Controller (CC) comes in. The detection patterns registered in each BSG are transmitted back to the CC who determines which of the BSG's in each triple of BSGs has succeeded. That is, a BSG typically needs detection of two photons to herald success. Those detections might actually occur in completely different nodes, $B$ and $F$ for example -- so node $A$ needs to be told by the CC which BSGs succeeded in order to block/unblock outputs appropriately.     

\subsubsection{Step 3 - Fusion and single qubit measurement}

The architecture iterates. Fusion of the small entangled states into larger ones is a probabilistic process and can be passively multiplexed. This is not always necessary - there are variants of photonic architectures where percolation theory guarantees that we have a state universal for quantum computing straight after the fusion. Either way, the final fate of all modes is to be measured, typically in the $X$ or $Z$ basis. At this stage our qubit is highly delocalized and it is a key feature of the architecture that the appropriate single qubit measurement can be performed locally within each sub-node, and results transmitted to the CC who can collate them so as to work out which outcome obtained.

 \subsection{More details}\label{sec:moredetails}
 
 We now discuss in a bit more detail how one choice of universal probabilistic procedures - entangled stage generation, fusion and single logical qubit measurement, work under passive multiplexing. There are more complicated versions of all of these procedures which are more efficient (in the sense of both using less resources and having higher success probabilities), but we will focus on the easiest versions for pedagogical clarity. 
 
 \subsubsection{Logical Qubit encoding}
 
 In Paper I the architecture was built on encoding the logical qubits in any one of a large number of orthogonal states. In particular, a logical qubit was defined across $2m$ modes, with the logical qubit state $\ket{0}$ corresponding to a single photon in \emph{any one} of the first $m$ modes (with vacuum in the second $m$ modes). Conversely, the logical qubit state $\ket{1}$ corresponds to a single photon in \emph{any one} of the second $m$ modes (with vacuum in the first $m$ modes). For example, with $m=4$ we could have: 
 \begin{eqnarray}\label{eq:multirailencode}
	\ket{0}&\Leftrightarrow  & \fket{0,0,1,0}\fket{0,0,0,0}=:\fket{1_3}\fket{0} \nonumber \\ 
	\ket{1}&\Leftrightarrow  & \fket{0,0,0,0}\fket{0,1,0,0}=:\fket{0}\fket{1_2}.
\end{eqnarray} 

\begin{figure}
\includegraphics[scale=0.55, clip=true,bb=5.0cm 13.2cm 11.8cm 23.7cm]{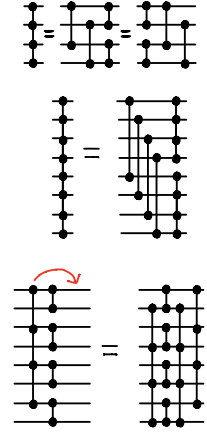}
\caption{Useful, easily generalized, matrix identities. Unlike previous figures where this notation loosely encompassed a variety of suitable interferometers, from now on we restrict to $H^{\otimes k}$ type Hadamard interferometers.}
\label{fig:Hidentities}
\end{figure}

 For simplicity, from now on we take $m=2^k$ and only use interferometers described by $H^{\otimes k}$. Such interferometers obey a bunch of simple recursive identities (see examples in Fig.~\ref{fig:Hidentities}) that are useful in terms of designing procedures for the DNA which implement desired logic.
 
 The DNA uses logical\footnote{Again: ``Logical'' in this paper never refers to a logical qubit from the theory of fault tolerance. It refers to the `bare' qubit for which a single qubit basis state might correspond to many physically-distinct states. Fault tolerant logical qubits need to be built on top of this architecture, e.g. via FBQC\cite{bartolucci2021fusionbased}} states that are `uniformly spread' versions of the ones of Paper I. In particular, DNA logical qubits are defined by taking the original multirail logical states and passing the $2m$ modes through an interferometer described by $H^{\otimes k} \oplus H^{\otimes k}$:   
\begin{align}\label{eq:logicalencoding}
	\tket{0} &{\Leftrightarrow}& {\sum_{i=1}^m h_i^{(j)}\fket{1_i}} &\otimes  {\fket{0,\ldots,0}} &=:\tfket{1_j}\otimes \tfket{0} \notag\\ 
	\tket{1} &{\Leftrightarrow}&  {\fket{0,\ldots,0}}          &\otimes {\sum_{i=1}^m h_i^{(j')}\fket{1_i}} &=:\tfket{0}\otimes \tfket{1_{j'}} 
\end{align}
 Here $j,j' \in \{ 1,\ldots m\}$ labels the mode that the single photon was in prior to passing through the interferometer, and the coefficients $h^{(j)}_i$ are the $(i,j)$ entries of $H^{\otimes k}$, which are all of the form $\pm\frac{1}{\sqrt{2^k}}$. 
 
 The fact that in general $j\neq j'$, i.e. that we can use many ``microscopically distinct'' physical states to encode the same logical state, is at the heart of the DNA. This feature - that there are many orthogonal states corresponding to a single logical qubit state - was also a key feature of of Paper I. In the DNA the orthogonality is carried by the relative phases between modes - all logical states have equal probability of finding the photon in a given mode if it is measured.

 \subsubsection{Bell state generation}
 \begin{figure}
\includegraphics[scale=0.55, clip=true,bb=3.0cm 5.7cm 18.5cm 24.3cm]{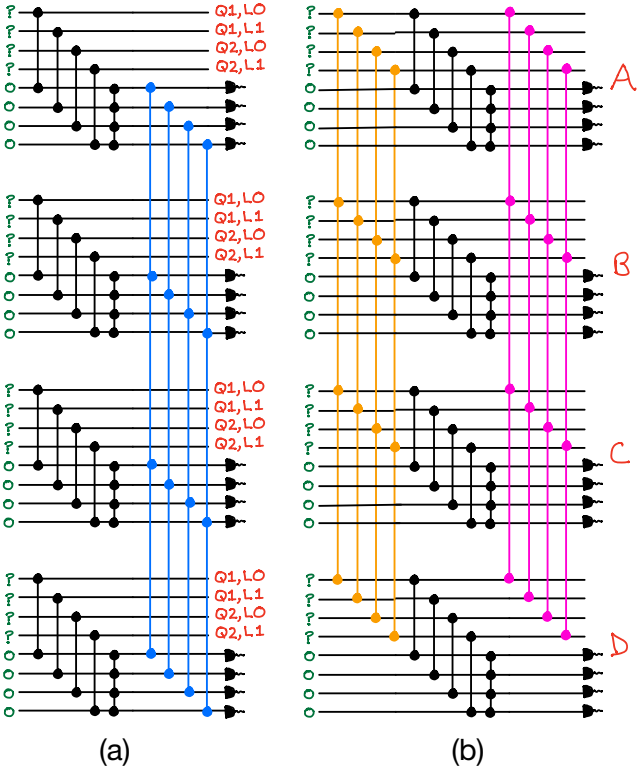}
\caption{Two interferometers that are identical (if vacuum is input on the modes indicated). The labelling on output modes of the form Q1L0 means ``Qubit 1, Logical 0'' etc.}
\label{fig:BSGmultirailfig}
\end{figure}

 The standard BSG takes four single photons along with four vacuum modes, and heralds detection of a Bell pair after two photons are detected. As discussed in detail in Paper I and depicted in Fig.~\ref{fig:BSGmultirailfig}(a), we can take any power-of-2 number such BSGs and add extra interferometers (blue in the figure) just prior to the detectors. Doing this now allows us to input the first single photon into \emph{any} of modes $1,9,17,25,$ (numbering top to bottom - i.e. any of the first ports of the separate BSGs)); the second single photon into \emph{any} of modes $2,10,18,26,$ and so on. When the procedure succeeds we may, for example, end up with a state like $\fket{1_9}\fket{1_{27}}+\fket{1_1}\fket{1_{20}}$ (the exact locations of the photons depends on which ports they were input and where the two were detected). In the terms of the original multirail encoding (see of Eq.\ref{eq:multirailencode}) this is a logical bell state $\ket{0}\ket{0}+\ket{1}\ket{1}$.  
 
 The circuit of Fig.~\ref{fig:BSGmultirailfig}(b) is identical to that of (a). Consider first the pink interferometers at the output. Their effect was to ``despread'' the output photons. They implement the (self inverse) transformation $H^{\otimes k}$ that maps between the logical $\ket{0}$ and $\tket{0}$ and $\ket{1}$ and $\tket{1}$ states. Thus if we remove the pink interferometers we obtain an output Bell state of the form $\tket{0}\tket{0}+\tket{1}\tket{1}$.
 
 We now identify the 4 BSGs as lying in nodes A,B,C,D of the DNA as described in the preceding section. Without the pink interferometers these nodes do not need to exchange quantum information. The orange interferometers now can be taken to be part of the passively multiplexed single photon source of Fig.~\ref{fig:singlephotonsource}(a). (For comparison, the top BSG in Fig.~\ref{fig:BSGmultirailfig}(b) could then be taken as the top BSG in Fig~\ref{fig:nodeAfigure}.)
 
 In summary we have seen that we can perform preparation of Bell states with flow of quantum and classical information corresponding to that claimed in Sections~\ref{sec:highleveloverview},\ref{sec:mediumleveloverview} such that we output a logical Bell state in the $\{\tket{0},\tket{1}\}$ basis.

As discussed in Paper I a key issue with encodings that allow many distinct orthogonal states to encode a single logical qubit, is that single qubit unitary operations are not automatically available. In particular, on a standard dual rail qubit the logical Hadamard is easy and deterministic (a beamsplitter), and so the ability to produce Bell pairs automatically implies the ability to produce the two-qubit cluster state by doing a Hadamard on one qubit. On a generic multirail encoded qubit the Hadamard can only be done with certainty using coherent switching, which we are trying to avoid. As in Paper I, however, we can use a modified Bell state generator that outputs states with a Hadamard operation ``pre-wired'' in.

Generalizations of standard state generation procedures for GHZ states and useful multi-qubit cluster states are all possible. It should be emphasized that the success probability of any of these procedures does \emph{not} depend on the number of nodes of the DNA (i.e. on the ``amount'' of the passive multiplexing), although depending on details the success probability may be reduced by a constant factor (meaning more copies of the process are required to be passively multiplexed in order to hit a target final success probability).

\subsubsection{Fusion}

Once we can generate entangled (cluster) states we want to be able to fuse them \cite{Browne2005}. Fusion is a effected via a parity measurement, which can be achieved by doing a partial Bell measurement. We want to be able to perform an operation that, for example, tells us the logical states of the qubits were ``the same'' without telling us which of the two options (both 0, both 1) actually pertains. 

\begin{figure}
\includegraphics[scale=0.55, clip=true,bb=4.5cm 18.4cm 13.5cm 24.3cm]{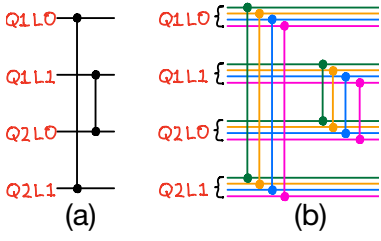}
\caption{ (a) Type-II fusion in standard dual rail encoding. Success is heralded by detecting one photon at one of the top or bottom modes, and the other photon in one of the middle modes. (Detectors not shown).   (b) Type-II fusion in a 4-node DNA. In the DNA the incoming logical states are uniformly spread over the four nodes (each node's interferometer is depicted with a different color). No interference is necessary between different nodes, they each apply a standard Type-II gate independently. Success is heralded by detecting one photon at one of the groups of top or bottom modes, and the other photon in one of the groups of the middle modes. (Detectors not shown)}
\label{fig:IIfusionfig}
\end{figure}

In the standard dual rail encoding of two qubits to do type-II fusion we apply beamsplitters between modes (1,4) and (2,3) (i.e. as in Fig.~\ref{fig:IIfusionfig}(a)). We see immediately that the $\ket{0}\ket{1}=\fket{1,0,0,1}$ and $\ket{1}\ket{0}=\fket{0,1,1,0}$ terms both lead to two photons detected in the same interferometer (both at (1,4) or both at (2,3)). However the $\ket{0}\ket{0}=\fket{1,0,1,0}$ and $\ket{1}\ket{1}=\fket{0,1,0,1}$ terms will always lead to one photon at modes (1,4) and one at (2,3). Moreover, the beamsplitter erases the information as to which mode the single photons originated from, and so we can only tell from the detection that the logical states of the qubits were the same, and not what they actually were individually. 

We want to achieve the same effect, but now with our delocalized logical states, which, in the notation of Eq.({\ref{eq:logicalencoding}}) take the general form:
\begin{align}
\tket{0}\tket{0}&=\tfket{1_j}\tfket{0}\tfket{1_k}\tfket{0}  \label{twologicalsA}\\
\tket{1}\tket{1}&=\tfket{0}\tfket{1_{j'}}\tfket{0}\tfket{1_{k'}} \label{twologicalsB}\\
\tket{0}\tket{1}&=\tfket{{1_j}}\tfket{0}\tfket{0}\tfket{{1_{k'}}} \label{twologicalsC}\\
\tket{1}\tket{0}&=\tfket{0}\tfket{{1_{j'}}}\tfket{{1_{k'}}}\tfket{0} \label{twologicalsD}
\end{align}

Consider Fig.~\ref{fig:IIfusionfig}(b), which depicts an example of two DNA qubits, with modes corresponding to different nodes but the same logical state grouped together. That is, normally we imagine all the green modes in one node of the DNA spatially separate from all the orange ones in another node and so on, but have now drawn them gathered together. If the first qubit is in the logical 0 state then it is in some quantum state with uniform amplitude aver the four modes labelled Q1L0, and similarly for the other logical states depicted. 

Imagine, for example, that after sending the incoming logical qubits through Fig.~\ref{fig:IIfusionfig}(b) the green node reports one photon in its first mode, and the pink node reports one photon in its fourth mode (the very top and bottom modes of Fig.~\ref{fig:IIfusionfig}(b)). Then this detection pattern can only have arisen from the logical state in Eq.~(\ref{twologicalsC}). This would be failure of the fusion gate. If, instead, the green node reported one photon in its first mode, while the orange node reports one photon in its second mode (the first and sixth modes of Fig.~\ref{fig:IIfusionfig}(b)) then this detection pattern arises with equal likelihood from the logical states in Eqs.~(\ref{twologicalsA}),(\ref{twologicalsB}), and we have succeded in fusing the logical states with even parity.

Similar types of reasoning can be used to show that Type-I and various boosted fusion protocols can all also be used directly ``as is'' within the DNA, implying that each node can operate as a standard level 0 quantum computer, the only difference being that the classical processing  of measurement results and instructions about what subsequent operations to perform must be made by the CC who, unlike the indivdual nodes, knows the logical state at any point in time.

\subsubsection{Single (logical) qubit measurement}

The final element we require to have universal quantum computation with our logical cluster states is the ability to do single (logical) qubit measurement. Measurement in the Z-basis $\{\tket{0},\tket{1}\}$ is simple because the photon is found in the first $m$ modes if the logical state is $\tket{0}$ and the the second $m$ modes if it is $\tket{1}$. Note also that unitary rotations around the logical z-axis are also easy, since rotating the phase of all the second $m$ modes of an encoded qubit we can readily perform $\tfket{1_j}\rightarrow e^{i\alpha}\tfket{1_j}$.

The key thing to show is that we can perform the logical Pauli $X$ measurement. In particular, if we single out a particular logical qubit of the computer then we can write the total state as 
\[\ket{A_0}\tfket{1_j}\tfket{0}+\ket{A_1}\tfket{0}\tfket{1_k}
\]
and we want to convince ourselves it is possible to collapse the state to $\ket{A_0}\pm\ket{A_1}$. Now if every node performs a beamsplitter on the two modes of the qubit they hold then we obtain this evolution:
\begin{align*}
&\ket{A_0}\left(\sum_ih^{(j)}_i\fket{1_i}\right)\tfket{0}+\ket{A_1}\tfket{0}\left(\sum_l h^{(k)}_l\fket{1_l}\right) \\
\rightarrow&\ket{A_0}\left(\sum_ih^{(j)}_i(\fket{1_i}\fket{0}+\fket{0}\fket{1_i})\right)\\
+&\ket{A_1}\left(\sum_l h^{(k)}_l(\fket{0}\fket{1_l}-\fket{1_l}\fket{0})\right)
\end{align*}
If all nodes then measure their pair of modes, and the photon is found in node number $s$, the state will collapse as desired to 
\[
h^{(j)}_s\ket{A_0}\pm h^{(k)}_s\ket{A_1},
\]
with the $\pm$ dependent upon whether $x\in\{1,\ldots,m\}$ or $x\in\{m+1,\ldots,2m\}$. This gives us our logical $X$ measurement.

As outlined briefly in Paper I, although the basis the single logical qubits are measured in needs to be adapted on the fly, it is in principle possible to still perform them without using a coherent switch; that is, using only a blocking switch as was used in the rest of the architecture. (But it would be pretty silly to do, since you lose a lot of loss tolerance as you can no longer tell the difference between a qubit lost and a qubit absorbed by the measurement blocking switch!)

\subsection{Further observations}

\subsubsection{Some phases important, others not?}

It can be a bit confusing at first to see why relative phases between photons that head off to different nodes (e.g. the $h^{(j)}_i$) are important, but relative phases between the nodes themselves are not. (For example in Fig.~\ref{fig:singlephotonsource}(b) the optical fibers would not need to be kept phase stable with respect to each other). A quick way to see why is to consider that by the end of the computation all photons will have been counted - we will know how many ended up in each node.  Thus we will have postselected ourselves into a part of the quantum state with $n_A$ photons at node $A$, $n_B$ at node $B$ and so on. In such a state if there are unknown random phases of the form $e^{in_A\theta_A}e^{i n_B\theta_B}..$ they all factor out completely. 

\subsubsection{A mode is a mode is a mode}

The DNA has been described in terms of generic modes, and although spatial modes are the most natural implementation to have in mind when trying to envision the structural flow of quantum and classical information, the whole architecture works just as well with any types of modes. As in Paper I, using time/frequency multirail modes is very natural for many of the types of heralded sources currently being built in integrated photonics. In particular, both of these architectures make heavy use of transformations (such as $H^{\otimes k}$ or a discrete fourier transform) which erase information, and single photon detectors are fast enough these days that they would (implicitly) perform such erasure on photons of different frequencies\cite{fastdetector1,fastdetector2}. 

\subsubsection{Bosons are overrated}

As described above described the number of nodes can be chosen to ensure near-deterministic, passively multiplexed single photon sources. However we saw that the logical operations did not care about the number of nodes. So, one can imagine instead that we use a very large number of nodes, so that at the end of the day each node has (with exponential likelihood) at most one photon. This limit is interesting as a way of understanding\cite{aliens} the limits of entanglement distribution by single photons. But it also makes clear that ultimately all we need is interference - interpreted as erasure of undesired quantum information - and as such non-bosonic schemes could conceivably make use of the ideas of Paper I and the present paper. 

\subsubsection{A crazy variant - don't ever touch the quanta!}

A different and somewhat extreme variant of the DNA shows that it is, in principle, possible to build a photonic quantum computer such that the single photon modes at most only pass through a single switch, namely the switch just prior to the adaptive single logical qubit measurement which chooses the algorithm we are running. That is, unlike the scheme laid out in the main text, the photons do not pass any blocking or coherent switches at all, unless they happen to be part of the final cluster state and undergo an adaptive measurement.

 This somewhat extreme variant is to take the possibility of using delays outlined in Fig.~\ref{fig:singlephotonsource}(c) seriously through all probabilistic elements and levels of the quantum computer. For example, consider building Bell pairs. We inject 4 single photons into the first BSG. At this step we do not even produce the four single photons that may potentially be required for a second attempt at making a Bell pair in the second BSG. That is, we are blocking the pump laser to all the other single photons which may or may not be required for creating this particular Bell pair. [Other passively multiplexed Bell pairs are being created elsewhere simultaneously. For example,  referring back to Fig.~\ref{fig:nodeAfigure}, each grouping of three BSG's would be staggered temporally in a manner similar to Fig.~\ref{fig:singlephotonsource}(c).] 
 
  If the first BSG succeeds then great, we never unblock the single photons for the other BSGs. If it does not, then we unblock the four photons for a second attempt in the second BSG. Note that the single photons already operate on a slow clock rate because they are passively multiplexed as per Fig.~\ref{fig:singlephotonsource}(c). Note also that even if the first BSG succeeds, its outputs must enter a very long delay to ensure that regardless of which BSG it is that ultimately succeeds, passive multiplexing is able to proceed. As we progress through the different levels of the quantum computer we have to keep increasing our delay lines so that we can always go back to turning the single photon sources off at the pump. 
  
  There are methods described in Paper I which allow one to `coherently erase' time-bin information that perhaps could ameliorate the rapid blow up of delay lines for this variant. (These should not be confused with the frequency/time+detectors possibility mentioned previously, which concerns the much shorter timescales within a single pulse).

\begin{acknowledgments}
Eric Johnston (EJ) realized the extreme architectural variant in the very last section was possible. He also wrote the excellent photonic simulation tools used to confirm the results in these papers. It was Damien Bonneau's questions and insights about phase (non)-stability in Paper I which prompted the investigation yielding the results of Paper II. Thanks to Jake Bulmer and EJ for comments on the drafts. As always the whole Quantum Architecture and System Architecture team at PsiQuantum are an endless source of inspiration and motivation. 
\end{acknowledgments}

\bibliography{LOQCpapers5.bib}

\end{document}